\documentstyle[12pt]{article}
\begin{document}
\def\thefootnote{\fnsymbol{footnote}}
\begin{flushright}
KANAZAWA-94-17  \\
October, 1994
\end{flushright}
\vspace{ .7cm}

\begin{center}
{\LARGE\bf Neutron Electric Dipole Moment under Non-Universal
Soft SUSY Breaking Terms}\\
\vspace{2cm}
{\Large Tatsuo Kobayashi}
\footnote[1]{e-mail:kobayasi@hep.s.kanazawa-u.ac.jp},~~
{\Large Masahiko Konmura}
\footnote[2]{e-mail:konmura@hep.s.kanazawa-u.ac.jp},~~
{\Large  Daijiro Suematsu}
\footnote[3]{e-mail:suematsu@hep.s.kanazawa-u.ac.jp}
\vspace{3mm} \\
{\Large  Kiyonori Yamada}
\footnote[4]{e-mail:yamada@hep.s.kanazawa-u.ac.jp} ~and~
{\Large  Yoshio Yamagishi}
\footnote[5]{e-mail:yamagisi@hep.s.kanazawa-u.ac.jp}
\vspace {1cm}\\

{\it Department of Physics, Kanazawa University,\\
        Kanazawa 920-11, Japan}
\end{center}
\vspace{2cm}
{\Large\bf Abstract}\\
The electric dipole moment of the neutron (EDMN) is re-examined in a
general framework of the soft supersymmetry breaking.
We point out some features of the relation between the EDMN and
non-universal soft supersymmetry breaking terms.
We give the constraints on the soft scalar masses and the soft CP
phases, which
have the rather large dependence on the non-universality of the soft
breaking terms.
We also show that the soft CP phase $\phi_B$ which has no natural
suppression mechanism may not have large contribution to the EDMN
in a certain parameter region where the radiative symmetry breaking
occurs successfully.
$\phi_B$ may not need to be so small.

\newpage
\setcounter{footnote}{0}
\def\thefootnote{\arabic{footnote}}
The minimal supersymmetric standard model (MSSM) is now considered as
the most promising extension of the standard model (SM)\cite{n}.
Although the origin of the supersymmetry breaking is still left unknown
we can make various predictions by using the suitable parametrization of
its breaking.
This parametrization is known as the soft supersymmetry breaking terms.
Phenomenological features of the MSSM are determined by these soft
supersymmetry breaking parameters, which play a similar role to a
vacuum expectation value of the Higgs field in the SM.
Usually these soft supersymmetry breaking parameters are treated as the
universal ones as derived from a special type of supergravity theory.
The electric dipole moment of the neutron (EDMN) in the MSSM is also
studied under the assumption of the universal soft supersymmetry breaking
terms.
In such a study the EDMN is known to exceed the present
experimental bound $1.1 \times 10^{-25}$e\thinspace cm \cite{al}
unless either the CP phases in the soft breaking parameters are
unnaturally small of order $O(10^{-3})$
or the masses of superpartners of quarks are heavier than
$O(1)$TeV\cite{pw,dgh,egn}.
Both conditions seem not to be easily satisfied if we consider
the soft breaking terms on the basis of the naturalness in the general
way.
This means that the EDMN may be a very important phenomenon to study
the evidence of the supersymmetry and the origin of the supersymmetry
breaking.

As recently stressed, the soft supersymmetry breaking parameters are
non-universal in the effective theories which are derived from
the superstring theories and also the general supergravity
theories\cite{il,kl,bim}.
It is shown in some works that such a non-universality shows very
interesting effects in the gauge coupling unification, the radiative
symmetry breaking and so on\cite{ksy,mn}.
Very remarkable features which are not seen in the usual study
are found in those works.
The EDMN is usually very enhanced due to the presence
of superpartners and new CP phases in the supersymmetric models.
Therefore its re-examination under non-universal soft supersymmetry
breaking terms seems to be very important for the phenomenological study of
the supersymmetry and the supersymmetric model building.

In the present letter we investigate the EDMN in the MSSM with the
general soft supersymmetry breaking terms and discuss its relation to
the squark masses and the soft CP phases.
In the MSSM it is well-known that the EDMN has non-zero value at
the one-loop level due to the contributions of the gluinos, the charginos
and the neutralinos.
Here we concentrate ourselves only on the gluino contribution to see
the effects due to the non-universality of the soft supersymmetry
breaking terms to the EDMN.
For the full quantitative estimation of the EDMN we need to take
account of the chargino contribution.
We will comment on the chargino contribution later.

At first we briefly review the general formulae and new CP phases of the
soft supersymmetry breaking terms in the MSSM and then give an
explicit formula of the EDMN due to the gluino contribution.
The superpotential of the MSSM is written as
\begin{equation}
W=h^U_{IJ}Q^IH_2U^J +h^D_{IJ}Q^IH_1D^J +h^E_{IJ}L^IH_1E^J +\mu H_1H_2,
\end{equation}
where $I$ and $J$ are the generation indices.
The soft supersymmetry breaking terms are
\begin{eqnarray}
{\cal L}_{\rm soft}&=&-\sum_i \tilde m_i^2\vert z_i \vert^2
+(A^U_{IJ}h^U_{IJ}Q^IH_2U^J +A^D_{IJ}h^D_{IJ}Q^IH_1D^J
+A^E_{IJ}h^E_{IJ}L^IH_1E^J \nonumber \\
&+&B\mu H_1H_2+\sum_a{1 \over 2}M_a\bar\lambda_a\lambda_a + h.c.),
\end{eqnarray}
where the first term represents the mass terms of all the scalar
components in the MSSM.
In the last term $\lambda_a$ are the gaugino fields for the gauge group
specified by $a$.
The remaining terms are the scalar two and three points couplings.

Various works based on the superstring theories and also general
supergravity theories
suggest that these soft breaking parameters are generally
non-universal\cite{il,kl,bim}.
In general low energy effective supergravity theories are characterized
in terms of the K\"ahler potential $K$, the superpotential $W$ and
the gauge kinetic function $f_a$.
Each of these is a function of the ordinary chiral matter superfields
$\Psi^I$ and the gauge singlet fields $\Phi^i$ called moduli.
Usually it is assumed that the nonperturbative phenomena such as a gaugino
condensation occur in the hidden sector.
After integrating out the relevant fields to these phenomena,
the K\"ahler potential and the superpotential are expanded by the low
energy observable matter fields $\Psi^I$,
\begin{eqnarray}
&&K=\kappa^{-2}\hat K(\Phi,\bar \Phi)+
Z(\Phi,\bar \Phi)_{I{\bar J}}\Psi^I\bar \Psi^{\bar J}
+\left({1 \over 2}H(\Phi,\bar \Phi)_{IJ}\Psi^I\Psi^J+{\rm h.c.}\right)
+\cdots, \\
&&W=\hat W(\Phi)+{1 \over 2}\tilde \mu(\Phi)_{IJ}\Psi^I\Psi^J+
\tilde h(\Phi)_{IJK}\Psi^I\Psi^J\Psi^K +\cdots,
\end{eqnarray}
where $\kappa^2=8\pi/M_{\rm pl}^2$.
The ellipses stand for terms of higher orders in $\Psi^I$.
Using these functions the scalar potential $V$ can be written,
\begin{equation}
V=\kappa^{-2}e^G[G_\alpha(G^{-1})^{\alpha \bar \beta}G_{\bar \beta}
-3\kappa^{-2}]+({\rm D-term}),
\end{equation}
where $G=K+\kappa^{-2}\log \kappa^6 |W|^2$ and the indices $\alpha$ and
$\beta$ denote $\Psi^I$ as well as $\Phi^i$.
The gravitino mass $m_{3/2}$ which characterizes the scale of the supersymmetry
breaking is expressed as
\begin{equation}
m_{3/2}=\kappa^2e^{\hat K}|\hat W|.
\end{equation}
In order to get the low energy effective theory from eq.(5) we take the flat
limit
$M_{\rm pl}\rightarrow \infty$ with $m_{3/2}$ fixed.
Through this procedure we get the superpotential (1) and the soft
supersymmetry breaking terms (2).
In the superpotential Yukawa couplings are rescaled as
$h_{IJK}= e^{\hat K/2}\tilde h_{IJK}$ and $\mu$ term is effectively
expressed as
\begin{equation}
\mu =e^{\hat K/2}\tilde \mu +m_{3/2}H -F^{\bar j}\partial_{\bar j}H,
\end{equation}
where $\mu$ should be understood as $\mu_{H_1H_2}$.
{}From this expression we can find that the favorable $\mu$ scale of order
$m_{3/2}$ can be remarkably induced.
However, it should be noted that the scale of $\mu$ crucially depends
on its origin and a case such as $|\mu|/m_{3/2} \ll 1$ can also occur.
This case will be interesting to consider the effects of the soft CP
phases on the EDMN.
Each soft breaking term is expressed by using $K$ and $W$ as
follows\cite{kl},
\begin{eqnarray}
&&\tilde m_{I{\bar J}}^2=m_{3/2}^2Z_{I{\bar J}}
-F^i{\bar F}^{\bar j}[\partial_i\partial_{\bar j}Z_{I{\bar J}}
-(\partial_{\bar j}Z_{K{\bar J}})Z^{K{\bar L}}(\partial_iZ_{I{\bar L}})]
+\kappa^2V_0Z_{I{\bar J}}, \\
&&A_{IJ}=F^i[(\partial_i +{1 \over 2}{\hat K}_i)h _{IJK}
-Z^{\bar ML}\partial_iZ_{\bar M(I} h_{JK)L}]/h_{IJK}, \\
&&B=F^i[(\partial_i +{1 \over 2}{\hat K}_i)\mu
-Z^{\bar ML}\partial_iZ_{\bar M(I} \mu_{J)L}]/\mu
-m_{3/2} \nonumber \\
&&\qquad +[F^i(\partial_i +{1 \over 2}{\hat K}_i)F^{\bar j}
-2m_{3/2}F^{\bar j}]\partial_{\bar j}H/\mu,
\end{eqnarray}
where $F^i$ are F-terms of $\Phi^i$ and $\partial_i$ denote
$\partial/\partial \Phi^i$.
$V_0$ in eq.(8) is the cosmological constant
expressed as
$V_0=\kappa^{-2}(F^i{\bar F}^{\bar j} \partial_i \partial_{\bar j}
\hat K-3m_{3/2}^2).$\footnote{In these formulae we do not assume
that the cosmological constant vanishes.}
Requiring the cosmological constant to be zero or sufficiently small,
we get $\vert F^i\vert=O(m_{3/2})$.
{}From this we find that the soft breaking terms $m_{I\bar J}$, $A_{IJ}$
and $B$ are generally non-universal but characterized by the gravitino
mass $m_{3/2}$.
The gaugino masses $M_a$ are derived through the following
equation,
\begin{equation}
M_a={1 \over 2}{\bar F}^{\bar j}\partial_{\bar j}\log {\rm Re} f_a,
\end{equation}
where the subscript $a$ represents a gauge group.
This shows that $M_a$ is also characterized by $m_{3/2}$.

The values of these soft breaking terms are at most $O(m_{3/2})$.
However, it should be noted that this does not mean the nonexistence of
the large difference among the soft breaking parameters.
In some models the soft breaking terms with the different order of
magnitudes are given\cite{bim,ksyy}.
These soft breaking terms are considered as the values at $M_{\rm pl}$.
The non-universality at high energy scale does not necessarily
mean the non-universality at the low energy when the quantum
corrections are taken into account.
It is also well-known that this non-universality for the scalar masses
at the low energy region is strictly constrained by the FCNC
phenomena\cite{gm}.
In the following study we concentrate ourselves to the case where
the low energy non-universality of the soft scalar masses remains without
contradicting the FCNC constraints.
As such a typical example, we consider the soft scalar masses in which
the up and down sector squark mass matrices have the similar form and
$\tilde m_{U_R}^2=\tilde m_{D_R}^2 \not=\tilde m_{Q_L}^2$.\footnote{
This type of soft scalar mass is discussed in ref.\cite{ksy}, where the
interesting feature of the gauge coupling unification is pointed out.}
In addition we assume that $A_{IJ}^U=A_U$ and $A_{IJ}^D=A_D$ for simplicity.

The soft breaking parameters $A_f$, $B$ and $M_a$ are generally complex
and then become new origins of the CP violation which do not exist in
the SM.
All of the phases of these parameters are known not to be physically
independent.
We can extract the physically independent phases from them in the
usual way\cite{dgh}.
We make the VEVs of the Higgs fields $H_1$ and $H_2$ real by the
appropriate redefinition of $H_1$ and $H_2$ so as to make $B$ real.
If we note that the complex phases in the gaugino masses are common
in eq.(11), we can make the gaugino mass real by the use of the
R-transformation and summarize the new CP phases associated to the soft
breaking terms in the following form,
\begin{equation}
\phi_{A_f}={\rm arg}(A_fM^\ast), \quad \phi_B={\rm arg}(BM^\ast).
\end{equation}
The non-universality of $A_f$ terms introduces the CP phase for each
Yukawa coupling.
These new CP phases can cause the new contributions to the EDMN.

Now we proceed to express the formula of the EDMN explicitly
keeping the non-universality of the soft breaking terms\cite{egn}.
As mentioned earlier, we only consider the gluino contribution whose
Feynmann diagram is shown in Fig.1.
To calculate this diagram we need an explicit form of a squark mass
matrix ${\cal M}_f^2$.
For the $f$-type squark $(f=U, D)$ it is explicitly written down as
\begin{equation}
\left( \begin{array}{cc}
\vert m_f\vert^2 +\tilde m_{fL}^2
+m_Z^2\cos2\beta(T_f^3-Q_f\sin^2\theta_W) & m_f(A_f+R_f\mu^\ast) \\
m_f^\ast(A_f^\ast+R_f\mu) & \vert m_f\vert^2 +\tilde m_{fR}^2
+m_Z^2\cos2\beta Q_f\sin^2\theta_W
\end{array}\right),
\end{equation}
where $m_f$, $\tilde m_{fL}$ and $\tilde m_{fR}$ are masses of the
$f$-quark, the corresponding left-handed squark and the right-handed
squark, respectively.
$T_f^3$ is the third component of the weak isospin of left-handed quark
$f$ and $Q_f$ is an electric charge of the quark $f$.
$R_f$ is defined by using
$\tan\beta\equiv\langle H_2\rangle / \langle H_1\rangle$ as
\begin{equation}
R_f=\left\{
\begin{array}{cc}
\cot\beta & ( {\rm for}~~ T_f^3=\displaystyle{1 \over 2}),\\
 \noalign{\vskip 0.2cm}
    \tan\beta & ( {\rm for}~~ T_f^3=-\displaystyle{1 \over 2}).
\end{array}\right .
\end{equation}
Although ${\cal M}_f^2$ is a 6 $\times$ 6 matrix, here we extract the part
corresponding to the first generation to estimate the EDMN.
This treatment will be justified because the generation mixing
off-diagonal components of ${\cal M}_f^2$ should be suppressed from the FCNC
constraints.

The contribution to the EDM of a quark $f$ from the diagram in Fig.1 is
\begin{equation}
d_f^g/e={\alpha_S \over 3\pi}\sum_{i=1}^2{\rm Im}(S_{2i}S_{1i}^\ast)
{Q_f \over m_g}r_i\int_0^1dx{x(1-x) \over 1-x+r_i-x(1-x)s_i}
\end{equation}
where $r_i= m_g^2/\tilde m_i^2$, $s_i=m_f^2/\tilde m_i^2$ and
$m_g$ is the gluino mass.
$S_{ij}$ is the element of the unitary matrix $S$ which diagonalizes the
squark mass matrix ${\cal M}_f^2$ as ${\cal M}_{\rm diag}^2
=S^\dagger {\cal M}_f^2S$.
The eigenvalues of this matrix are represented as $\tilde m_i^2$.
After evaluating these matrix elements $S_{ij}$, we get the final form
of $d_f^g$ as,
\begin{equation}
d_f^g/e={\alpha_S \over 3\pi}{Q_f\sin\gamma_f \over m_g}
\left[{4Z^2_f \over Y^2_f+4Z^2_f}\right]^{1 \over 2}
\left[r_2I(r_2)-r_1I(r_1)\right]
\end{equation}
where $\tilde m_1^2$ and $\tilde m_2^2$ are mass eigenvalues of $M_f^2$.
They are explicitly written down as
\begin{eqnarray}
&&\tilde m_{1f}^2={m_g^2 \over 2}\left[ X_f+\sqrt{Y^2_f+4Z^2_f} \right],
\nonumber \\
&&\tilde m_{2f}^2={m_g^2 \over 2}\left[ X_f-\sqrt{Y^2_f+4Z^2_f} \right].
\end{eqnarray}
Here the parameters $X_f, Y_f$ and $Z_f$ are defined as
\begin{eqnarray}
&&\displaystyle{X_f={\tilde m_{fL}^2 \over m_g^2}+{\tilde m_{fR}^2
\over m_g^2}
+{m_Z^2 \over m_g^2}\cos2\beta T_f^3}, \nonumber \\
&&\displaystyle{Y_f={\tilde m_{fL}^2 \over m_g^2}-{\tilde m_{fR}^2
\over m_g^2}
+{m_Z^2 \over m_g^2}\cos2\beta (T_f^3-2Q_f\sin^2\theta_W)}, \nonumber \\
&&\displaystyle{Z_f={1 \over m_g^2} \vert m_f(A_f +R_f\mu^\ast)\vert}.
\end{eqnarray}
In the derivation of these formulae we use the fact $m_f \ll m_g$ for
$f=U, D$.
The function $I(r)$ has the following form
\begin{equation}
I(r)={1 \over 2(1-r)^2}\left[1+r+{2r\ln r \over 1-r} \right].
\end{equation}
In this expression $s_i = m_f^2/\tilde m_i^2$ is neglected
because the quark mass is small enough compared to the soft breaking scalar
masses.
In eq.(16) an angle $\gamma_f$ can be written down as\footnote{
Here we assume that the arg($m_f$) is small enough not
to bring the strong CP problem.}
\begin{equation}
\tan\gamma_f= {
\vert A_f \vert \sin\phi_{A_f} +\vert R_f\mu^\ast\vert \sin\phi_B
 \over
\vert A_f \vert \cos\phi_{A_f} +\vert R_f\mu^\ast\vert \cos\phi_B}.
\end{equation}
In the case that $\phi_{A_f}$ and $\phi_B$ are sufficiently
small this reduces to the usually known form,
\begin{equation}
\gamma_f \sim
{\vert A_f \vert \over \vert A_f\vert +\vert R_f\mu^\ast\vert}
\phi_{A_f}
+{\vert R_f\mu^\ast \vert \over \vert A_f\vert +\vert R_f\mu^\ast\vert}
\phi_B.
\end{equation}
To reconstruct the EDMN from ones of the quarks, we follow the conventional
method and use the result of the nonrelativistic quark model
\begin{equation}
d_n={1 \over 3}(4d_d-d_u).
\end{equation}

Now we analyze the EDMN using these formulae in the non-universal
soft breaking case.
In this analysis we assume $\gamma_U =\gamma_D\equiv \gamma$ and
$A_U=A_D$, for simplicity.
To see the effects of the non-universality in the $u$- and $d$-
squark masses on the EDMN we plot the contour lines of
$d_n^g/e\sin\gamma$ in the $(\tilde m_R/m_g)$-$(\tilde m_L/m_g)$ plane
in Figs.2 $\sim$ 4.
Each graph corresponds to the various settings of
$m_g$ and $\vert A_f+ R_f\mu^\ast\vert$ values.
Here it should be noted that in these figures the contours are drawn for
$d_n^g/e\sin\gamma$ but not for the direct values of $d_n^g/e$.
Thus for the comparison of the present results to the experimental
bound we must estimate $\sin\gamma$.
As is easily seen from eq.(20), $\sin\gamma$ is of order one as far as
both of $\tan\phi_{A_f}$ and $\tan\phi_B$ are $O(1)$.
This is independent of the values of $\vert A_f\vert$ and
$\vert R_f\mu^\ast\vert$.
In this case Figs.2 $\sim$ 4 directly represent the value of the EDMN
which can be compared to the experimental bound.
If both of $\phi_{A_f}$ and $\phi_B$ are less than $O(1)$,
$\gamma$ will be approximately expressed by eq.(21) and
$\gamma =O(\phi_{A_f}), O(\phi_B)$.
Taking account of these, we can read off the conditions to satisfy the
experimental bound of the EDMN from these figures.
The constraints usually quoted in the universal soft breaking case seem
to be rather weakened by various combined effects of the non-universal
soft breaking parameters.
Furthermore a suitable combination of the non-universality may present
the interesting possibility for the EDMN.
In particular, as seen from eq.(21),
as far as $\vert A_f\vert$ and $\vert R_f\mu^\ast\vert$ are not the
same order either $\phi_{A_f}$ or $\phi_B$ will mainly contribute
to the EDMN.
This may open the new possibility for the solution of the soft
CP phases as seen later.

Here we comment on the rather large dependence of the EDMN on
$\tilde m_L, \tilde m_R, m_g$ and $\vert A_f +R_f\mu^\ast\vert$.
The large values of $\tilde m_L$ or $\tilde m_R$ are required to
suppress the EDMN.
However, it should be remarkable that if the only one of them is
sufficiently heavy the EDMN can be largely suppressed
as expected from the consideration for the squark mass eigenvalues.
As $\vert A_f +R_f\mu^\ast\vert$ increases $d_n^g/e\sin\gamma$
proportionally does.
This feature is easily understood if we note that
$\vert A_f +R_f\mu^\ast\vert$ characterize the left-right mixing of
the squark mass matrix.
The value of $d_n^g/e\sin\gamma$ increases according to the
decrease of the gluino mass if we keep $\tilde m_R/m_g$ and
$\tilde m_L/m_g$ constant.
Following to the usual RGE study, the soft masses such as
$m_g \gg \tilde m_L, \tilde m_R$
seem to be difficult to realize at the low energy region.
The large gluino mass will make the squark masses the same order
as the gluino mass through the renormalization effect.\footnote{
In the present study we consider only the gluino contribution and thus
the non-universality of the gaugino masses is irrelevant.
However, when we estimate the chargino contribution it should be also
taken into account.}
The reasonable region such as $m_g {^<_\sim} \tilde m_L, \tilde m_R$
has the tendency to make the EDMN small.
Anyway the non-universality of the soft supersymmetry breaking terms
can fairly affect the value of the EDMN.

Next we study the necessity of the natural suppression of the soft
 CP phases.
As was shown in the previous part, $\sin\gamma$ should be small enough
not to exceed the experimental bound of the EDMN if all squark masses
are $O(100)$~GeV.
This is usually considered to be equivalent to the condition that
both of the soft CP phases $\phi_{A_f}$ and $\phi_B$ are less than
$10^{-2} \sim 10^{-3}$ depending on $m_g$ and
$\vert A_f+R_f\mu^\ast\vert$.
{}From the viewpoint of the naturalness such small phases seem not to
be expected in the general soft supersymmetry breaking schemes.
In the following parts we shall propose a natural explanation for this
problem.

Recently it is suggested that the phase $\phi_{A_f}$ can be small enough
not to contradict with the EDMN bound in the models derived from the
superstring theories with the supersymmetry breaking due to the F-terms
of a dilaton and moduli.
In ref.\cite{bim} it was shown that the dilaton dominated supersymmetry
breaking
suppresses the phase $\phi_{A_f}$ sufficiently.
This is because the phase structure of $A_f$ and $M_a$ in eqs.(9) and
(11) has a certain similarity.
On the other hand Choi pointed out in ref.\cite{c} that the various
complex phases contributing $\phi_{A_f}$ are tuned to the value less
than $O(10^{-3})$ by the dynamical mechanism based on the Peccei-Quinn
 symmetry on the dilaton and moduli.
Unfortunately there is no such general suppression mechanisms for
the phase $\phi_B$.
The origin of $\mu$ term is not determined as shown in eq.(7) and the
structure of $\phi_B$ completely depends on its origin as seen from eq.(10).
It seems very difficult to suppress $\phi_B$ naturally.\footnote{
An only known mechanism to suppress $\phi_B$ naturally is to replace
the $\mu$-term by a Yukawa coupling of a singlet field and Higgs fields.
In this model $\phi_B$ is reduced to the $\phi_{A}$ type phase and then
$\phi_B$ will also be small automatically\cite{bim}.}
However, the existence of natural suppression mechanisms of $\phi_{A_f}$
can open the new possibility of the sufficient suppression of the EDMN.
Instead of finding the suppression mechanism of $\phi_B$,
it seems more promising to investigate this new possibility that the
EDMN may be sufficiently suppressed even if the phase $\phi_B$ is
not so small.

For this purpose we will consider the case of
$\vert A_f\vert \gg \vert R_f\mu \vert$.
We assume that the smallness of $\phi_{A_f}$ is guaranteed by the
above mentioned mechanism.
In such a case the value of $\gamma_f$ can be estimated as
\begin{equation}
\gamma_f \sim
\phi_{A_f} +{\vert R_f\mu^\ast \vert \over \vert A_f\vert }\sin\phi_B.
\end{equation}
The contribution from $\phi_B$ can be suppressed by a
factor $\vert R_f\mu^\ast \vert/ \vert A_f\vert $ even if
$\phi_B$ is $O(1)$.
The main issue of this scenario is the consistency between the
radiative symmetry breaking and the smallness of
$\vert R_f\mu^\ast \vert/ \vert A_f\vert $.
Using eqs.(16) and (23), we can estimate the contribution to the EDMN from
$\phi_B$ as
\begin{equation}
d_n^g/e\sin\phi_B = {1 \over 3}{\vert \mu \vert \over \vert A\vert}
(-X_U\cot\beta+4X_D\tan\beta)
\sim {4 \over 3}{\vert \mu\vert \over \vert A\vert}X_D\tan\beta
\end{equation}
where $A_f$ is assumed as $A_U=A_D=A$.
The approximate value of $4X_D/3$ can be read off from Figs.2 $\sim$ 4.
Generally $\tan\beta >1$ is expected in the radiative symmetry
breaking scenario due to the large top Yukawa coupling.
Thus the second similarity in eq.(24) is deduced.
As far as $\tan\beta\sim O(1)$ and the masses of all superpartners are
$\sim$100~GeV, the necessary condition to satisfy the experimental
bound of the EDMN is estimated as $\vert\mu\vert /\vert A\vert< 10^{-2}
\sim 10^{-3}$.
As suggested by the previous argument on the soft breaking terms,
$A$ is expected as $O(m_{3/2})$ where the magnitude of $m_{3/2}$ is
dependent on the supersymmetry breaking mechanism.
On the other hand the scale of $\mu$ depends on its
origin and then $\vert\mu\vert/\vert A\vert<10^{-2} \sim 10^{-3}$
may be possible.
However, the small $\mu$ may yield the light chargino and conflict the
experimental constraint.
The chargino mass matrix has the following form
\begin{equation}
\left( \begin{array}{cc}
\mu & \sqrt 2m_W\sin\beta \\
\sqrt 2m_W\cos\beta & M_2 \\
\end{array} \right)
\end{equation}
and then the squared mass eigenvalues of the charginos are
\begin{equation}
m_{\omega_{1,2}}^2=
{1 \over 2}\left[\mu^2 +M_2^2+2m_W^2\sin 2\beta
\mp \sqrt{(\mu^2-M_2^2)^2+4m_W^2(\mu +M_2)^2\sin 2\beta}\right].
\end{equation}
If we consider the region where both of $\mu$ and $M_2$ are
sufficiently small and also $\tan\beta$ is not so larger than one,
the chargino mass does not conflict the
present experimental bound 45~GeV.

Based on these considerations we concentrate ourselves to the
parameter region such as
\begin{equation}
B, A, \tilde m_i \gg \mu, M_2
\end{equation}
at the low energy region.
In this region we practice the RGE study to examine the radiative
$\rm SU(2)\times U(1)$ breaking and estimate the top quark mass.
As is well known, the small $\mu$ increases
the value of $\tan\beta$ significantly.
This effect may cancel the smallness of $\mu$ and make our scenario
less attractive.
To avoid such situation,
we take $B(M_{\rm pl})$ larger than other soft parameters.
This is because $\tan\beta$ does not depend on $\mu$ directory
but depends on $B\mu$.
At the tree level analysis, $\beta$ is expressed as\cite{inou}
\begin{equation}
\sin 2\beta = \frac{2B\mu}{m_{H_1}^2+m_{H_2}^2+2\mu^2}.
\end{equation}
{}From this one can find that
the value of $m_{H_1}$ also influences $\tan\beta$
in the same way as $B$
\footnote{In the most case of the radiative symmetry breaking,
$\vert m_{H_2}^2+\mu^2\vert$ is
about $O(10^{-1})(m_{H_1}^2+\mu^2)$ so that we need not to take account
of its effect.}.
The non-universality of soft scalar masses may be applied to
not only squarks and sleptons but also the Higgs sector.\footnote{
{}From the viewpoint of radiative $\rm SU(2)\times U(1)$ breaking,
it is interesting to vary initial values
of Higgs masses from other scalar ones\cite{mn}.}
By choosing $m_{H_1}(M_{\rm pl})$ smaller than
$m_{H_2}(M_{\rm pl})$,
we can reduce $\tan\beta$ furthermore.
The non-universality between $\tilde m_U$ and $\tilde m_D$ also affects
the running of Higgs masses through RGEs and
one can expect similar effects as above.
However, such effects are indirect and negligible unless we
assume extremely large squark mass hierarchy,
which often causes color SU(3) breaking.
Combining these effects we can find a suitable parameter region on the
basis of RGEs study.
As such an example we list a typical set of soft supersymmetry breaking
parameters at $m_Z$,
$$ \vert \mu\vert/\vert A\vert =2.9\times 10^{-2}, \qquad
\tan\beta =2.6. $$
For these values the top quark mass becomes 148~GeV and the lighter
chargino mass is $\sim 45$~GeV.
The constraint on $\phi_B$ seems to be largely reduced
to the rather natural value as $\phi_B \sim O(10^{-1})$ for these
parameters.
The combined effects of the non-universality of the soft supersymmetry
breaking parameters are expected to weaken the constraints from the
EDMN furthermore.

Finally we should comment on the chargino contribution.
In this case the $\phi_A$ dependence
is largely suppressed due to the small Yukawa couplings.
The chargino contribution mainly comes from the $\phi_B$ dependent
part of the $d$-quark EDM and is expressed as\cite{egn}
\begin{eqnarray}
d^c_d/e\sin\phi_B &{^<_\sim}& {\alpha_{em} \over 4\pi\sin^2\theta_W}
{M_2\vert \mu\vert \over m_{\omega_2}^2-m_{\omega_1}^2}
{m_f \over \tilde m_i^2}\times O(1) \nonumber \\
&\sim& 5.1\times 10^{-25}\left({1~{\rm TeV} \over \tilde m_i}\right)^2
\left({m_f \over 10~{\rm MeV}}\right)
{M_2\vert \mu\vert \over m_{\omega_2}^2-m_{\omega_1}^2}
\times O(1),
\end{eqnarray}
In the parameter region presented above as an example
this chargino contribution is expected to be sufficiently within the
experimental bound for $\phi_B\sim 10^{-1}$ even if $m_i \sim
O(100)$~GeV.
This is because the appearance of the additional suppression factor
$${M_2\vert \mu\vert \over m_{\omega_2}^2-m_{\omega_1}^2} \sim
{\vert\mu\vert \over 2m_W\sin^{1/2}2\beta} \sim 7.5\times 10^{-2}$$
for the above parameters.
The present experimental bound of the EDMN may be
reconciled with the soft supersymmetry breaking without introducing
unnatural assumptions on the parameters.
Although the present parameter region does not seem so wide,
it may inform us the suitable non-universality of
the soft breaking parameters in the MSSM.

In summary, we re-examined the EDMN under the general soft supersymmetry
breaking parameters.
We pointed out some features of the relations between the EDMN and soft
supersymmetry breaking parameters which seems not to be mentioned
explicitly before in the universal soft breaking framework.
We also showed that the soft CP phase $\phi_B$ whose natural
suppression mechanism is not known up to now does not have large
contribution to the EDMN in the certain parameter space where the radiative
symmetry breaking occurs successfully.
This may be an interesting non-universal parameter region of the MSSM.
We may not need to require that $\phi_B$ is so small.
FCNC constrains severely the soft masses of the squarks, in particular,
 with the same charge.
It requires their degeneracy at $m_Z$ scale.
On the other hand the study of the EDMN may give us some other
knowledge for the squark masses as is shown in this paper.
If we combine these, we may get a useful insight for the whole
structure of the soft squark masses.
Moreover, the EDMN may give us useful information of the soft breaking
parameters in the MSSM.
{}From this point of view the more precise theoretical study of the EDMN
seems to be very important.
Also the improvement of the experimental bound of the EDMN is strongly
desired.

\vspace{5mm}
\noindent
{\Large\bf Acknowledgement}

\vspace{2mm}
One of the authors(T.~K.) would like to thank C.~S.~Lim for the useful
discussions on the CP-violation.
The work of T.~K. is partially supported by Soryuushi Shogakukai and
the work of D.~S. is supported in part by a Grant-in-Aid for Scientific
Research from the Ministry of Education, Science and Culture
(\#05640337 and \#06220201).

\newpage

\newpage
{\large\bf Figure Captions}
\vspace{3mm}\\
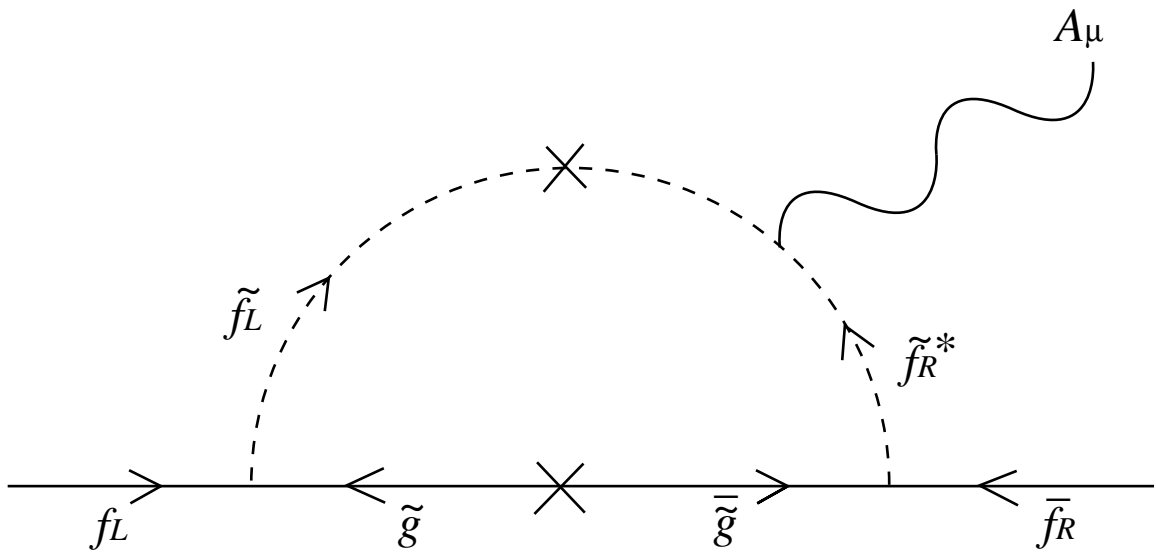
{\bf Fig.1}\\
A Feynmann diagram of the gluino contribution to the EDMN.
\vspace{3mm}\\
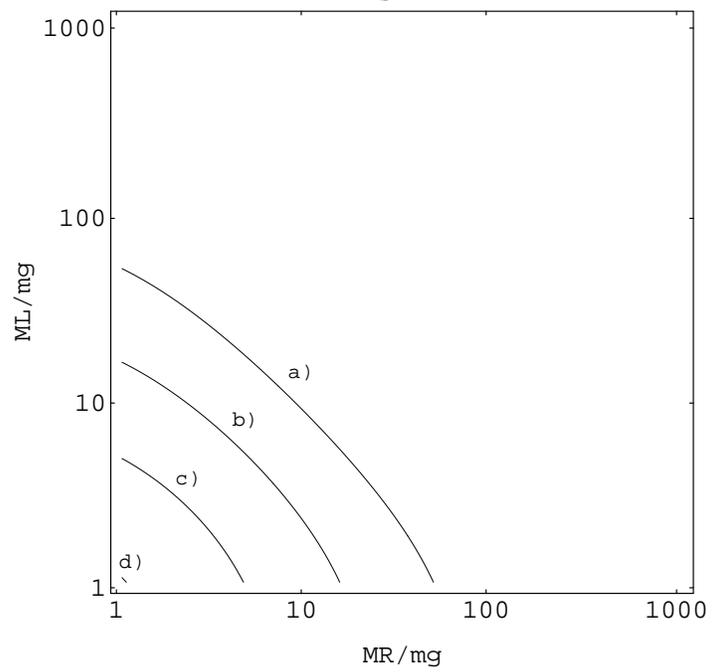
{\bf Fig.2}\\
The contours of $d_n^g/e\sin\gamma$ in the $(\tilde m_R/m_g)$-$(\tilde
m_L/m_g)$ plane at $m_g=100$~GeV.
$\vert A_f+R_f\mu^\ast\vert$ is chosen as 100~GeV and 1000~GeV in
Fig.2A and Fig.2B, respectively.
Each contour corresponds to
a)$2.0\times 10^{-26}$ cm,
b)$2.0\times 10^{-25}$ cm,
c)$2.0\times 10^{-24}$ cm,
d)$2.0\times 10^{-23}$ cm and
e)$2.0\times 10^{-22}$ cm.
\vspace{3mm}\\
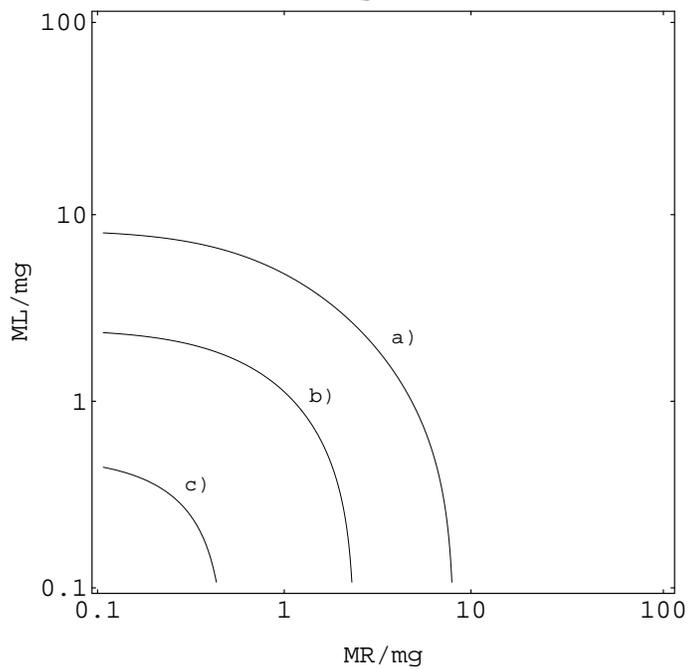
{\bf Fig.3}\\
The contours of $d_n^g/e\sin\gamma$ in the $(\tilde m_R/m_g)$-$(\tilde
m_L/m_g)$ plane
at $m_g=500$~GeV.
The setting of $\vert A_f+R_f\mu^\ast\vert$ is the same as Fig.2.
Each contour represents the same value as Fig.2.
\vspace{3mm}\\
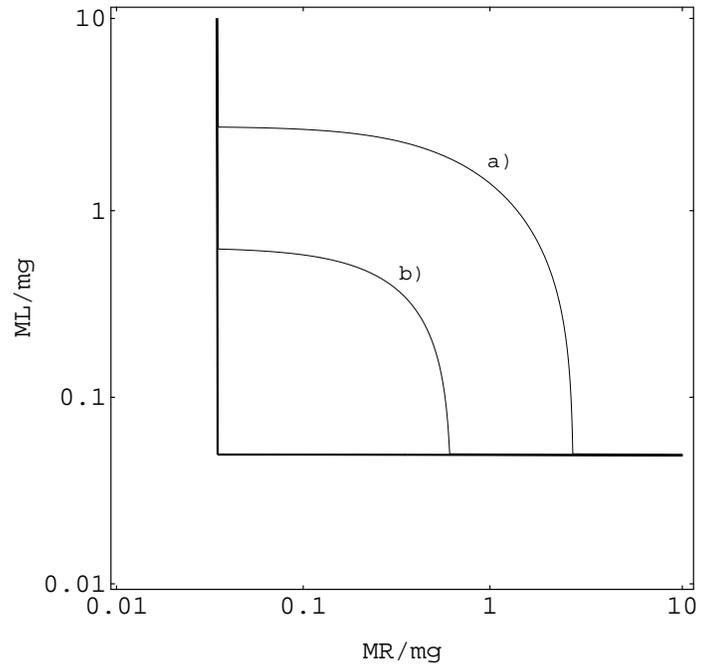
{\bf Fig.4}\\
The contours of $d_n^g/e\sin\gamma$ in the $(\tilde m_R/m_g)$-$(\tilde
m_L/m_g)$ plane at $m_g=1000$~GeV.
The setting of $\vert A_f+R_f\mu^\ast\vert$ is the same as Fig.2.
Each contour represents the same value as Fig.2.
The region outside the vertical and horizontal lines is prohibited
because the squared mass eigenvalues of squark become negative.

\begin{thebibliography}{99}
\bibitem{n}For a review, see for example, H.-P.~Nilles, Phys. Rep.
{\bf C110}(1984)1 and references therein.

\bibitem{al}I.~S.~Altarev at al., Phys. Lett. {\bf B276}(1992)242.

\bibitem{pw}J.~Ellis, S.~Ferrara and D.~V.~Nanopoulos, Phys. Lett.
{\bf 114B}(1982)231,

W.~Buchm\"uller and D.~Wyler, Phys. Lett, {\bf 121B}(1983)321,

J.~Polochinski and M.~B.~Wise, Phys. Lett.
{\bf 125B}(1983)393.

\bibitem{dgh}M.~Dugan, B.~Grinstein and L.~Hall, Nucl. Phys.
{\bf B255}(1985)413.

\bibitem{egn}G.~Ecker, W.~Grimus and H.~Neufeld, Nucl. Phys. {\bf B229}
(1983)421,

Y.~Kizukuri and N.~Oshimo, Phys. Rev.{D46}(1992)3025.

\bibitem{il}L.~E.~Ib\'a\~nez and D.~L\"ust, Nucl. Phys.
{\bf B382}(1992)305.

\bibitem{kl}V.~S.~Kaplunovsky and J.~Louis,
Phys. Lett. {\bf B306}(1993)269.

\bibitem{bim}A.~Brignole, L.E.Ib\'{a}\~{n}ez and C.Mu\~{n}oz,
Nucl. Phys. {\bf B422}(1994)125.

\bibitem{ksy}T.~Kobayashi, D.~Suematsu and Y.~Yamagishi, Phys. Lett.
{\bf B329}(1994)27.

\bibitem{mn}A.~Lleyda and C.~Mu\~noz, Phys. Lett. {\bf B317}(1993)82,

N.~Polonsky and A.~Pomarol, preprint UPR-0616-T(1994)(hep-ph9406224),

D.~Matalliotakis and H.~P.~Nilles, preprint TUM-HEP-201/94
(hep-ph/9407251)

M.~Olechowski and S.~Pokorski, preprint MPI-PHT/94-40(hep-ph/9407404),

D.~Choudhury, F.~Eberlein, A.~K\"onig, J.~Louis and S.~Pokorski,
preprint MPI-PhT/94-51(hep-ph/9408275).

\bibitem{ksyy}T.~Kobayashi, D.~Suematsu, K.~Yamada and Y.~Yamagishi,
preprint KANAZAWA-94-16 (hep-ph/9408322).

\bibitem{gm}F.~Gabbiani and A.~Masiero, Nucl. Phys. {\bf B322}(1989)235,

J.~Hagelin, S.~Kelley and T.~Tanaka, Nucl. Phys. {\bf B415}(1994)293.

\bibitem{c}K.~Choi, Phys. Rev. Lett. {\bf 72}(1994)1592.

\bibitem{inou}L.~Ib\'a\~nez and G.~Ross, Phys. Lett. {\bf 110B}
(1982)215,

K.~Inoue, A.~Kakuto, H.~Komatsu and S.~Takeshita,
Prog. Theo. Phys. {\bf 68}(1982)927,

L.~Alvarez-Gaume, J.~Polchinski and M.~B.~Wise, Nucl.
Phys. {\bf B221}(1983)495,

J.~Ellis, D.~V.~Nanopoulos and K.~Tamvakis, Phys. Lett. {\bf B121}(1983)
123.

\end{thebibliography}
\end{document}